\documentclass[preprint,showpacs,amsmath,amssymb]{revtex4-1}
\usepackage{mathrsfs}
\usepackage{slashed}

\usepackage{graphicx,color,multirow}
\usepackage{dcolumn}
\usepackage{bm}
\newcommand{\nn}{\nonumber}

\newcommand {\beq} {\begin{equation}}
\newcommand {\eeq} {\end{equation}}
\newcommand {\bea} {\begin{eqnarray}}
\newcommand {\eea} {\end{eqnarray}}
\newcommand{\GeV}{{\rm\ GeV}}

\newcommand{\rmnum}[1]{\romannumeral #1}

\begin{document}

%-------------------------------------------------------------------------------------------------------

\title{Searches for dark matter via charged Higgs pair production in the Inert Doublet Model at $\gamma\gamma$ collider}

\author{Yang Guo-He$^a$}%\email{yusihe234@163.com}
\author{Song Mao$^a$}\email{songmao@mail.ustc.edu.cn}
\author{Li Gang$^a$}%\email{lig2008@mail.ustc.edu.cn}
\author{Zhang Yu$^{b,a}$}%\email{dayu@ahu.edu.cn}
\author{Guo Jian-You$^a$}%\email{jianyou@ahu.edu.cn}

\affiliation{$^a$ School of Physics and Material Science, Anhui University, Hefei 230601, China}
\affiliation{$^b$ Institutes of Physical Science and Information Technology, Anhui University, Hefei 230601, China}
\date{\today}

%-------------------------------------------------------------------------------------------------------
\begin{abstract}
The Inert Doublet Model(IDM) is one of the simplest extensions beyond Standard Model(SM) with an extended
scalar sector, which provide a scalar dark matter particle candidate. In this paper, we investigate the double charged Higgs production at $\gamma\gamma$ collider. By scanning the whole parameter space, we obtain the parameter points corresponding to the correct relic abundance of dark matter. After applying all theoretical and experimental constraints, the parameter space for the existence of dark matter is extremely restricted. We perform the analysis for the signal of $H^+H^-$ production in the IDM and the SM backgrounds, and the optimized selection conditions are chosen in kinematic variables to maximize signal significance. Comparing signal with backgrounds, we obtain the parameter points which can be detected at future $\gamma\gamma$ collider experiments.

\end{abstract}

\maketitle

\section{Introduction}
\par
After the discovery of the Higgs boson \cite{higgs1,higgs2}, the Standard Model (SM) of particle physics has achieved great success in describing the
particles up to energies of about 1 TeV \cite{Baak:2012kk}.  However, there are still many questions as well as many unexplained phenomena remain, such as,
the symmetry of matter and antimatter, the sources of CP violation, the nature of dark matter (DM) particle. All these mean that the Standard Model is perhaps just a low energy approximation of a more fundamental theory. At the same time, the Standard Model of Big Bang Cosmology, known as "$\Lambda$CDM", is successful
in describing the Universe large scale structure formation and evolution, the state of the early Universe and the abundance of the different form of matter and energy \cite{Spergel:2003cb, DelPopolo:2008mr, DelPopolo:2013qba}, whose predictions are supported by new observation ($e.g.$, lensing of the CMB \cite{Smith:2007rg, das}, B-mode polarisation \cite{Hanson:2013hsb}, the kinetic Sunyaev Zeldovich(SZ) effect). The astrophysical and cosmological observational evidences have confirmed the existence of DM and provided the DM density in the universe \cite{Bertone04}. However, the Standard Model of particle physics could not provide enough dark matter. Until now, we have little information about the properties of dark matter particles.
Among all the DM candidates, Weakly Interacting Massive Particles (WIMPs) is a promising option.
Since it offers the DM candidates to interpret the relic abundance naturally in rebuilding the thermal history of the universe \cite{Feng2008a}.

Among various extended scenarios beyond SM, the Inert Doublet Model (IDM) is one of the simplest models to explain the WIMP dark matter. In this model, an isospin doublet scalar field is added to the SM Higgs sector, which is assumed to be odd under a discrete $Z_2$ symmetry. After electroweak symmetry breaking, four $Z_2$ odd scalar particles are generated, $i.e.$, one CP-even $H$, one CP-odd $A$ and two charged $H^\pm$ scalar bosons. Among them, the lightest scalar boson may serve as a dark matter candidate. The $Z_2$ symmetry ensures that these new scalar particles can not decay into final states only including the SM particles. In addition, the additional isospin doublet scalar does not directly interact with the SM fermions at tree level. Their interactions with the Standard Model particles are achieved via gauge coupling and the quartic term with the SM Higgs in the scalar potential.

The lightest scalar Higgs in the IDM, as a dark matter candidate, needs to be able to reconstruct the correct DM relic abundance.
In Ref~\cite{chile_DM, Dolle:2009fn}, they found that three allowed mass regimes for the lightest Higgs satisfy the requirement of relic abundance.
The scalar dark matter particle has also been explored by various direct, indirect experiments and high energy colliders.
In current direct detection experiments, the dark matter mass has been constrained to be
around one half of the SM-like Higgs boson mass (125 GeV) or above about 500 GeV~\cite{chile_DM, IDM_DM_th3, Compressing_IDM}.
In Ref~\cite{Eiteneuer:2017hoh}, the authors investigate the constraint for IDM parameter space from dark matter annihilation induced
gamma-rays in dwarf spheroidal galaxies. The phenomenology for the IDM at hadron colliders has been studied in the literature, such as $H^+H^-$, $HH^{\pm}$, $HA$ pair production, followed by the subsequent decay chains $A \to ZH$, $H^{\pm} \to W^{\pm}H$ \cite{IDM_DM_th8, IDM_DM_th9, Compressing_IDM, LHC_direct_Cao, LHC_direct_Su, LHC_run-I_bound1, LHC_run-I_bound2,song_ctp,Ahriche:2018ger}.
The prospects for discovery of scalar dark matter particle in the IDM at future lepton colliders has been discussed~\cite{IDM_ILC1, IDM_ILC2, Kalinowski:2018ylg}. Moreover, the constraint for IDM using vector boson fusion is also investigated in Ref.~\cite{Dercks:2018wch}. With the option of an $e^+e^-$ collider, it also can be run in $\gamma\gamma$ mode (at an energy scale similar to that of the primary electron-positron design). The charged Higgs pair can product directly in the IDM at $\gamma\gamma$ collider mode. Compared with $e^+e^-$, $\gamma\gamma$ collider can provide higher cross section in the high energy region because the charged Higgs pair is mainly dominated by s-channel diagrams at the tree-level in $e^+e^-$ collider. In this paper, we will investigate the production of charged Higgs pair in $\gamma\gamma$ collider.

The paper is organized as follows. In Section II we briefly describe the framework of inert doublet model.
In Section III we calculate the relic abundance of dark matter in the IDM.
In Section IV, we summarize all the theoretical and phenomenological constraints on the scalar potential of the IDM.
In Section V, we present the numerical results of the total and differential cross sections for the charged Higgs pair production.
In Sec VI, we analyse the charged Higgs pair signatures at $\gamma\gamma$ collider with its subsequent decay $H^{\pm} \to W^{\pm}H$.
Finally, a short summary is given in Section VII.

\section{The Inert Higgs Doublet Model}

The inert doublet model is one of the simplest extension of Standard Model (SM), which contain two SU(2) complex scalar fields $\Phi_{1}$ and $\Phi_{2}$, which are invariant in discrete $Z_2$ symmetry. The scalar field $\Phi_{1}$ is almost the same as the SM Higgs field, which is $Z_2$ even with hypercharge $Y=1$. Under $Z_2$ symmetry, $\Phi_{1}$ satisfy the transformation $\Phi_{1} \rightarrow  \Phi_{1}$.
The field $\Phi_2^{}$ is odd under the $Z_2^{}$ symmetry with hypercharge $Y=1/2$, which satisfy the transformation $\Phi_{2} \rightarrow  - \Phi_{2}$ under $Z_2$.
Under the electroweak symmetry $SU(2)_{L}\times U(1)_{Y}$ and the discrete $Z_2$ symmetry, the Higgs sector potential of the IDM is
\begin{align}
	V(\Phi_1,\Phi_2) =&\mu^2_1|\Phi_1|^2+\mu^2_2|\Phi_2|^2
			+\frac{1}{2}\lambda_1|\Phi_1|^4+\frac{1}{2}\lambda_2|\Phi_2|^4  \nonumber\\
			&+\lambda_3|\Phi_1|^2|\Phi_2|^2
			+\lambda_4|\Phi_1^\dagger\Phi_2|^2
			+\frac{1}{2}\{ \lambda_5(\Phi_1^\dagger\Phi_2)^2+h.c.\},   \label{eq:potential}
\end{align}
In the case of CP-conservation, all the parameters are real.
The theoretical constraints for these coupling parameters from perturbative unitarity are given in Refs.~\cite{Kanemura:1993hm,Akeroyd:2000wc}. In $Z_2$ symmetry, $\Phi_2$ has zero vacuum expectation value (VEV), and the SM like field, $\Phi_1$ takes part in the electroweak symmetry breaking (EWSB). After the EWSB, the doublet scalar fields are expanded around physical vacuum.
\begin{equation}
	\Phi_1^{}=
	\begin{pmatrix}
 		G^+\\
 		\frac{1}{\sqrt{2}}(h+v+i G^0)
	\end{pmatrix}
	,\  \Phi_2^{}=
	\begin{pmatrix}
 		H^+\\
 		\frac{1}{\sqrt{2}}(H+iA)
	\end{pmatrix},
\end{equation}
where $G^+$ and $G^0$ are the charged and neutral Goldstone bosons that are manifested as the longitudinal components of the gauge bosons, and $h$ is the SM-like Higgs boson with mass $m_h = 125$ GeV. The vacuum expectation value (VEV) of $\Phi_1$ is $v=246$ GeV.
The second doublet field $\Phi_2$  contain four $Z_2^{}$ odd scalar bosons, a CP even neutral scalar boson $H$,  a CP odd neutral scalar boson $A$, and two charged Higgs bosons $H^\pm$. After EWSB, the masses of these scalar bosons are given as
\begin{align}
	m_h^2&=\lambda_1 v^2, \label{eq:mhsq}\\
	m^2_{H^+}&=\mu_2^2+\frac{1}{2}\lambda_3 v^2, \label{eq:mHpsq}\\
	m_H^2&=\mu_2^2+\frac{1}{2}(\lambda_3+\lambda_4+\lambda_5)v^2, \label{eq:mHsq}\\
	m_A^2&=\mu_2^2+\frac{1}{2}(\lambda_3+\lambda_4-\lambda_5)v^2. \label{eq:mAsq}	
\end{align}
Assuming $\lambda_5<0$, the lightest CP even neutral scalar boson $H$ is stable, and could be a candidate of dark matter.
The IDM scalar sector can be specified by a total of six free parameters:
\begin{align}
\{ \lambda_1, ~\lambda_2, ~\lambda_3, ~\lambda_4, ~\lambda_5, ~\mu_2 \}.	
\end{align}
We introduce the useful abbreviations $\lambda_L = \frac{1}{2}(\lambda_3+\lambda_4+\lambda_5)$ and $\lambda_s = \frac{1}{2}(\lambda_3+\lambda_4-\lambda_5)$.  Through the above equations, the six parameters can be changed into a set of more meaningful parameters,
\begin{align}
\{ m_{H^{\pm}}, m_A, m_H, m_h, \lambda_L, \lambda_2 \},
\end{align}
where $m_{H^{\pm}}, m_A, m_H$ are the four $Z_2^{}$ odd scalar boson masses. $\lambda_L$ correspond to the coupling of the dark matter and SM-like Higgs boson, which is relevant for dark matter annihilation. The quartic coupling $\lambda_2$ correspond to self-interaction in the dark sector.

\vskip 5mm
\section{Thermal Relic abundance of dark matter}
The dark matter relic abundance is obtained by solving the non-equilibrium Boltzmann equation,
\begin{equation}
\frac { dn_{\chi} } {dt} + 3 \rm{H} n_{\chi} = - \left\langle \sigma \textit{v} \right\rangle  \left( n_{ \chi }^{2} - \left(n_{\chi}^{ eq } \right)^{2}, \right)
\end{equation}
where $n_{\chi}$ is the number density of dark matter particles, and $\rm{H}$ is the expansion rate of the universe, $\left\langle \sigma v \right\rangle$ is the thermally averaged annihilation cross section.
\par
By convention, we introduce the comoving number density $Y = n/s $, as well as substituting temperature for $x = \frac{m_{\chi} }{T}$, then
the derivative of $Y$ with respect to $x$ is
\begin{equation}
\frac { d Y } { d x } = - \frac {s} { \rm{H} x^{ 2 } } \left\langle \sigma v \right\rangle \left( Y^{ 2 } - \left( Y^{eq} \right)^{2} \right).
\end{equation}
Solving this equation, we can get the number density as the temperature. By integrating the function from $x = x_f$ to $x \to \infty$, we get the number density $Y_{\infty }$.
\par
Using the result $Y_{\infty }$, we get the final relic density $ \Omega h^{2} $,
\begin{equation}
\Omega h^{2} \equiv \frac {\rho_{\chi} } {\rho_{c} } h^{2} = \frac {m_{\chi} Y_{\infty} s_{\infty} } { 1.05 \times 10^{-5} \mathrm {GeV} \mathrm {cm}^{-3} },
\end{equation}
where  ${\rho_{c} } $ is the critical energy density of the universe, ${\rho_{\chi} } $  is the energy density of dark matter and $s_{\infty}$ is the entropy density in the present universe.

\par
We use the software Micromegas \cite{Belanger:2018ccd} to calculate the relic abundance of dark matter. The SM input relevant parameters are chosen as:
\begin{eqnarray}
&&m_{W}=80.379 \GeV, m_{Z}=91.1876 \GeV, m_{H}=125.18 \GeV, \nonumber \\
&&G_F=1.1663787\times 10^{-5}~{\rm GeV}^{-2}, m_{b}=4.18 \GeV, m_{t}=173.0 \GeV,\nonumber \\
&&\alpha_s(m_Z)=0.1181, \alpha=7.297352\times 10^{-3} , m_{c}=1.275 \GeV.
\end{eqnarray}
We choose ${m_{H}, m_{A}, m_{H^{\pm}}, \lambda_{L}, \lambda_{2}}$ as five independent input parameters of the IDM. The annihilation cross section is only calculated to the Leading Order (LO), thus the relic density is not sensitive to the parameter $\lambda_{2}$. If assuming the mass hierarchy $m_{H^{\pm} } \geq m_{A} > m_{H}$ for the inert scalar bosons, the lightest scalar $H$ will be a dark matter particle candidate. In general, the roles of $m_{H}$ and $m_{A}$ are interchangeable. Usually, its mass can be divided into three regions: \rmnum{1}.low mass (1-80\GeV), \rmnum{2}.intermediate mass (80-500\GeV), \rmnum{3}.high mass (500-1000\GeV).

Since $\lambda_2$ is only related to the self-coupling of inert particles, its variation has little effect on the relic abundance of dark matter particle, thus we fix it as $\lambda_2=0.01$. Then, we scan the other three mass parameters $m_{H}, m_{A}, m_{H^{\pm}}$ from 1\GeV~ to 1000\GeV~and $\lambda_L$ from $-0.75$ to 6.28.  When the DM relic density is in agreement with the Planck's measurements: $0.119 < \Omega_{DM}h^2 < 0.121$, these data points are saved.

\begin{figure}[t]
\begin{center}	
	\includegraphics[width=0.95\textwidth]{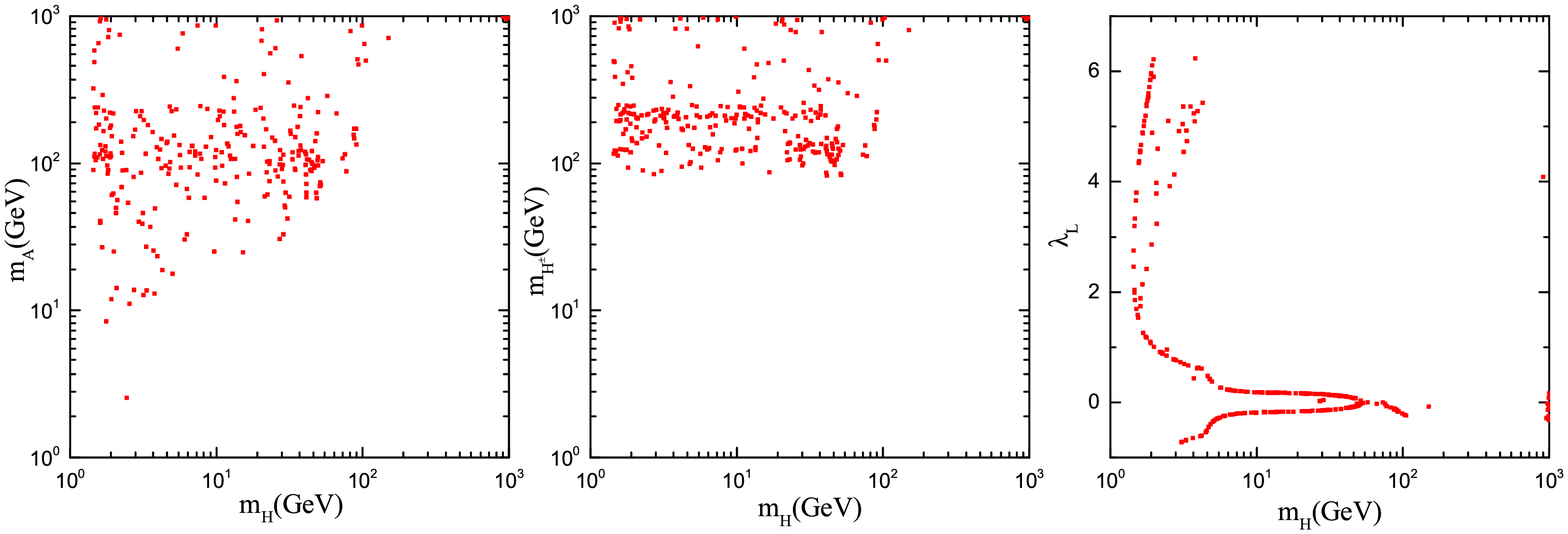}
	\caption{ \label{fig1} (color online) The projection of reserved points in $m_H-m_{A}$ (left), $m_H-m_{H^\pm}$ (middle), $m_H-\lambda_L $(right) plane. }
\end{center}
\end{figure}

We put these reserved points project onto two dimension plane in Fig.\ref{fig1}. The mass range of $ m_H$ is mainly concentrated on low mass region in the three figures, which below 150 GeV. The mass range of $m_{H^\pm}$ is mainly above 80 GeV. When $\lambda_L$ is close to zero, we can easily find a appropriate point in large $m_H$ area corresponding to correct relic abundance. When $\lambda_L$ is larger than 1, $m_H$ can only reserved in 1-3 GeV.

We select six groups of parameters as benchmark points, which are listed in the following:
\noindent
{BP1}: $\lambda_2$ = 0.01, $\lambda_L$ =-0.067137, $m_H$ = 48.47607, $m_{A} $=159.37519, $m_{H^\pm} $=171.83007\\
{BP2}: $\lambda_2$ = 0.01, $\lambda_L$ =-0.061830, $m_H$ =49.07796, $m_{A}  $=104.58929, $m_{H^\pm} $=177.79335 \\
{BP3}: $\lambda_2$ = 0.01, $\lambda_L$ = 4.37294, $m_H$ =1.57134, $m_{A}  $=183.29679, $m_{H^\pm} $= 192.24964 \\
{BP4}: $\lambda_2$ = 0.01, $\lambda_L$ = 0.13490, $m_H$ =32.23363, $m_{A} $=158.16124, $m_{H^\pm} $=194.83879\\
{BP5}: $\lambda_2$ = 0.01, $\lambda_L$ = -0.15581, $m_H$ =3.1979, $m_{A} $=196.59793, $m_{H^\pm} $=201.52571\\
{BP6}: $\lambda_2$ = 0.01, $\lambda_L$ = 5.83958, $m_H$ =1.87152, $m_{A} $=173.96496, $m_{H^\pm} $=186.41257\\

\begin{figure}[t]
\begin{center}	
	\includegraphics[width=0.49\textwidth]{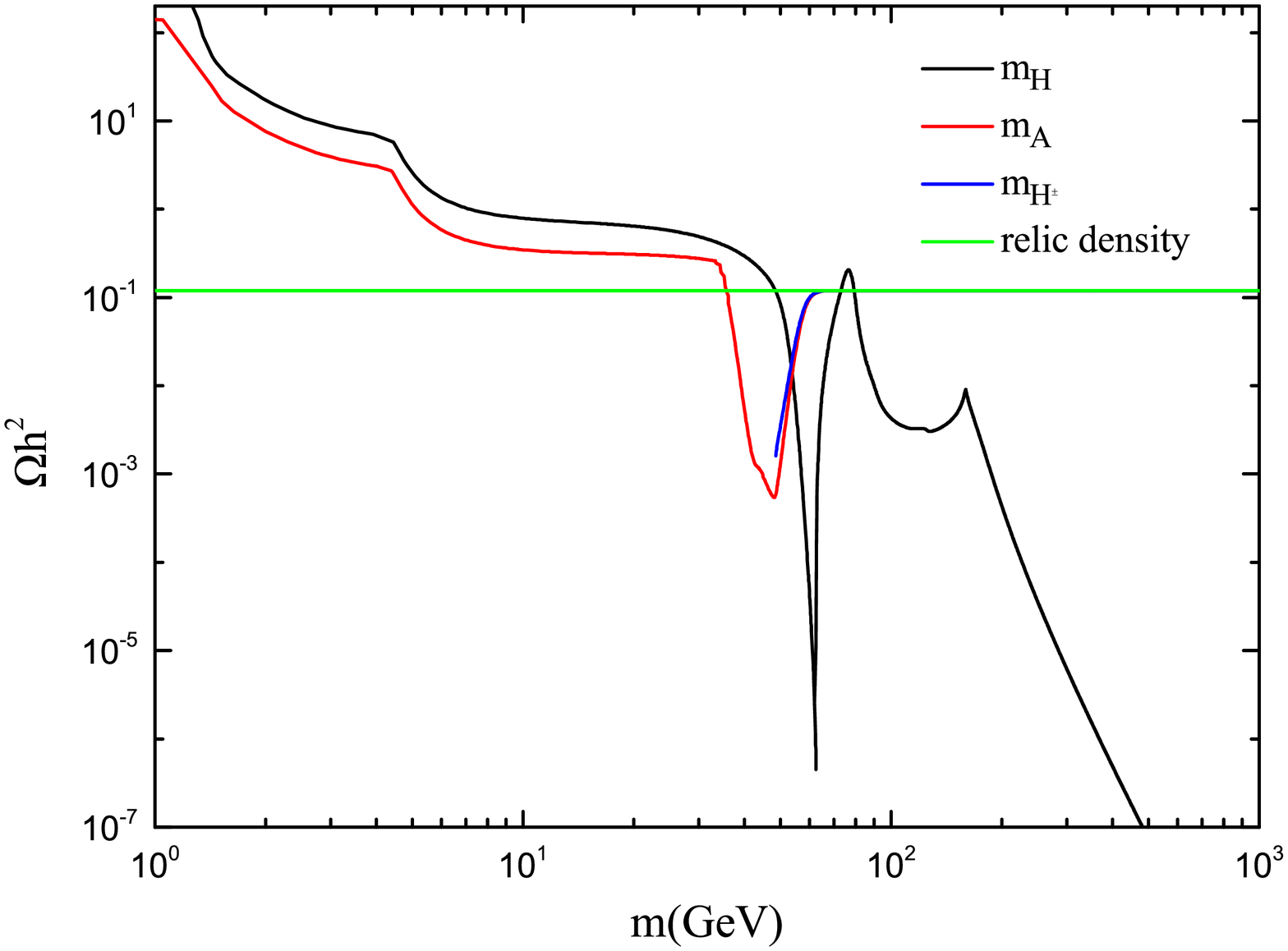}
	\includegraphics[width=0.49\textwidth]{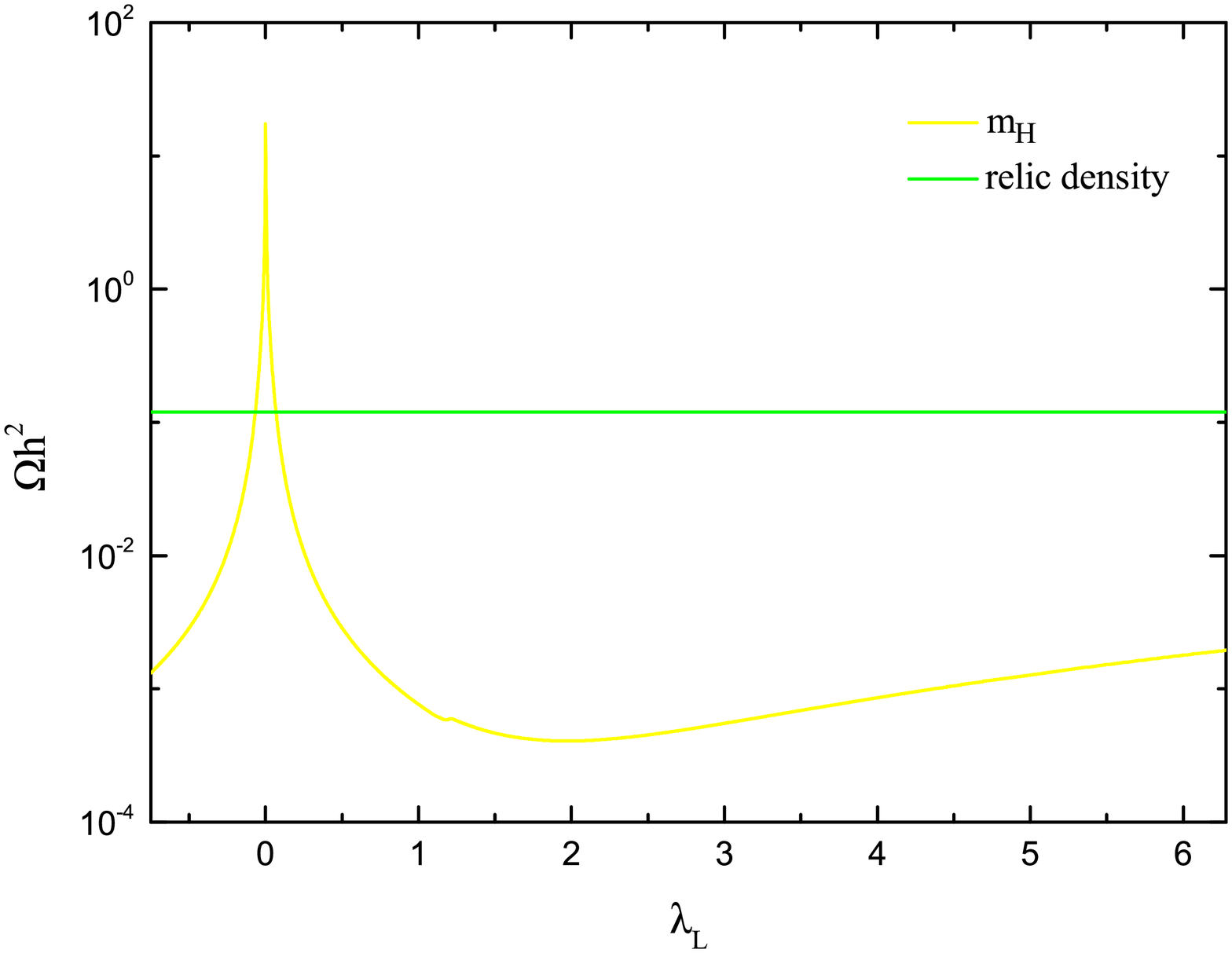}
	\caption{ \label{fig2} (color online) Relic density parameter $\Omega h^{ 2 }$ as a function of  $ m_H , m_{A}, m_{H^\pm}$ when other parameters are fixed (left). Relic density parameter $\Omega h^{2}$ as a function of  $\lambda_L$ when other parameters are fixed (right). }
\end{center}
\end{figure}

In Fig.\ref{fig2}, we present the function of the relic abundance of dark matter with the parameters $m_H$, $m_A$ $m_{H^\pm}$ and $\lambda_L$. Since all above six set of parameters have similar characteristics by drawing,
so we take BP1 as an example. From Fig.\ref{fig2}(left), we can see that, when the parameter $m_H$ or $m_A$ close to a half of higgs mass, the relic density is
greatly reduced. This is due to that these two particle are easily merged into a on-shell Higgs and
then decay into SM particles, thus the dark matter relic density has been dramatically reduced.
When $m_A$ or $m_{H^\pm}$ get into high mass region, they don't affect the relic density.
In Fig.\ref{fig2}(right), we provide the relic density $\Omega h^{2}$ as a function of $\lambda_L$ with the fixed other parameters. The relic density parameter
$\Omega h^{2}$ first increases and then decreases rapidly at the whole range of $\lambda_L$, and reaches its maximum value near zero. Even at some benchmark points, the point that corresponding to the correct relic abundance is not near zero, but the upward and downward trend always exists and reaches the maximum value near zero region.

\section{Constraints on the Model Parameters}

In this section, we summarize all the theoretical and experimental limitations for the extended scalar sector potential of the IDM.

First, the perturbation of the theory requires all the scalar coupling constants cannot exceed $4\pi$ \cite{GarciaCely:2014jha}.
\begin{align}
| \lambda_{1,2,3,4,5}| \leq 4\pi \,,\hspace{10pt}
|\lambda_3+\lambda_4\pm\lambda_5| < 4\pi \,,\hspace{10pt}
|\lambda_4\pm\lambda_5| < 8\pi \,,\hspace{10pt}
|\lambda_3+\lambda_4| < 4\pi.
\end{align}
In order to obtain a stable vacuum, the following parameters must be positive \cite{Gunion:2002zf,Gustafsson:2010zz,Khan:2015ipa},
\begin{align}
  \lambda_1^{} > 0,\ \lambda_2^{} > 0,\ \sqrt{\lambda_1^{} + \lambda_2^{}} + \lambda_3^{} > 0,\ \lambda_3^{}+\lambda_4^{} \pm |\lambda_5^{}| > 0. \label{eq:vs}
\end{align}

The unitary of the S-matrix for processes $2 \to 2$ scattering at the perturbative level requires all the couplings \cite{Ginzburg:2004vp,Branco:2011iw},

\begin{align}
|\lambda_3 \pm \lambda_4 |                                                                 		\leq  8 \pi, \quad
|\lambda_3 \pm \lambda_5 |                                                               		\leq   8 \pi, \quad
|\lambda_3+ 2 \lambda_4 \pm 3\lambda_5 |                                                       &\leq  8 \pi\nn\\
|-\lambda_1 - \lambda_2 \pm \sqrt{(\lambda_1 - \lambda_2)^2 + \lambda_4^2} |                       		& \leq  8 \pi\nn\\
|-3\lambda_1 - 3\lambda_2 \pm \sqrt{9(\lambda_1 - \lambda_2)^2 + (2\lambda_3 + \lambda_4)^2} |  & \leq  8 \pi\nn\\
|-\lambda_1 - \lambda_2 \pm \sqrt{(\lambda_1 - \lambda_2)^2 + \lambda_5^2} |                     		&\leq  8 \pi.\nn
\label{eq:unitary}
\end{align}

The Peskin-Takeuchi $S,\;T,\;U$ parameters are strictly limited by the electroweak precision observables. The deviation from the SM prediction $\Delta S$ and $\Delta T$ are experimentally given $\Delta S = 0.03\times0.09$, $\Delta T = 0.07\times0.08$. The contribution from the IDM can be calculated as in Ref.\cite{Arhrib:2012ia}. This typically prohibit large mass splittings among inert states, but for DM masses with $M_{H^0}\gtrsim 500$~GeV relatively small splittings are already required, especially when combined with the relic density constraint \cite{IDM_DM_th2}.

The experimental constraints for the inert scalars are mainly come from the large electron-positron collider (LEP) and large hadron collider(LHC) at CERN.

First, the constraints on the new scalar particles at LEP come from the measurements of the  $Z\rightarrow A H$, $Z\rightarrow H^+ H^-$, $W^\pm\rightarrow A H^\pm$ and $W^\pm\rightarrow H H^\pm$ decay, which imply that $M_{A^0} + M_{H^0} \geq M_{Z}$, $2M_{H^\pm}\geq M_{Z}$, $M_{H^\pm} + M_{H,A} \geq M_{W}$. Secondly, SUSY searches at at LEP\,II leads to constraints on the charged Higgs mass: the charged Higgs mass is constrained by $M_{H^\pm}\gtrsim 70$~GeV~\cite{Pierce:2007ut}, the bound on $M_{H}$ is also involved: if  $M_{H}<80$~GeV, then $|M_{A}-M_{H}| \le 8$ GeV, or else, $M_{A} \ge 110$~GeV\cite{LEP-II-H}.

The constraints on IDM at the LHC come mainly from the SM Higgs boson decay width. The new couplings from IDM can either increase the invisible branching ratio and/or alter the strength of the Higgs boson and diphoton coupling \cite{Arhrib:2012ia,Krawczyk:2013jta,Swiezewska:2012eh,Goudelis:2013uca}, which strictly limited the mass of the inert lightest scalar particle less than $M_h/2$, and has little restriction for the masses above $M_h/2$. Direct  di-leptons plus missing energy searches have also been performed to restrict the inert scalar masses in the region of $M_{H}\lesssim 60$~GeV and  $M_A \lesssim 150$~GeV \cite{Belanger:2015kga}.

From these constraints we find that the IDM is strongly restricted if the mass of inert scalar particles are less than 100 GeV and have little constraints for masses above 500\,GeV.

\vskip 5mm
\section{charged Higgs pair production at $\gamma\gamma$ collider}
\begin{figure}
\begin{center}
\includegraphics[width=0.85 \textwidth]{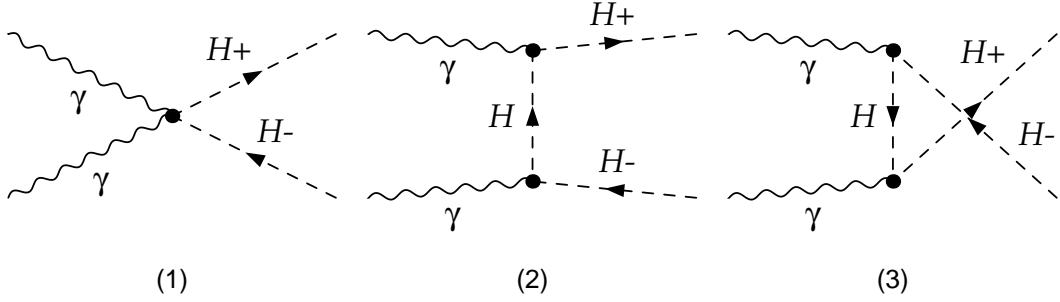}
\caption{ \label{born} The Feynman diagrams for process $\gamma \gamma \to H^+ H^-$ in IDM at $\gamma\gamma$ collider.}
\end{center}
\end{figure}

In order to maintain the symmetry of $Z_2$, the scalar particles in the IDM always produce in pairs at the collider. The lightest scalar particle $H$ in the IDM is stable and can be a dark matter candidate, other scalar particles will eventually decay into $H$ associated SM particles, such as $A \to H Z$, $H^{\pm} \to H W^{\pm}$. These scalar particles only couple to the Higgs boson and electroweak gauge bosons of the Standard Model, thus the production cross section for double scalar particles is usually small. However, the charged Higgs boson $H^\pm$ can couple to photons through electromagnetic interactions. Predictably, the cross section of the double charged Higgs production in $\gamma\gamma$ collider is considerable.
In this section, we consider the following process at $\gamma\gamma$ collider,
\begin{eqnarray}
\label{process}
\gamma \gamma \to H^+ H^-.
\end{eqnarray}
All the tree-level Feynman diagrams are presented in Fig.\ref{born}. The cross section of this process is only related to the mass of $H^{\pm}$ and is independent of the other four parameters $\lambda_{L}, \lambda_{2}, m_{H}, m_{A}$ in the IDM.

\par
The hard photon beam of the $\gamma\gamma$ collider can be obtained by using the laser backscattering technique at $e^+e^-$ linear collider \cite{Com1,Com2,Com3}. We denote $\hat{s}$ and s as the center-of-mass energies of the $\gamma\gamma$ and $e^+e^-$ systems, respectively. After calculating the cross section $\hat{\sigma}(\hat{s})$ for the subprocess
$\gamma\gamma \to  H^+ H^-$ in photon collision mode, the total cross section at an
$e^+e^-$ linear collider can be obtained by folding $\hat{\sigma}(\hat{s})$ with the photon distribution
function that is given in Ref.\cite{function,Ginzburg:1999wz}.
The cross section for the $e^+e^- \to \gamma\gamma \to H^+ H^-$ process is expressed as
\begin{equation}
\sigma_{tot}(e^+e^- \to \gamma\gamma \to H^+ H^-
,~s)=\int^{x_{max}}_{2m_H^{\pm}/\sqrt{s}} dz\frac{d{\cal
L}_{\gamma\gamma}}{dz} \hat{\sigma}(\hat{s}=z^2 s ).
 \end{equation}

The distribution function of photon luminosity is expressed as
\begin{eqnarray}
\frac{d{\cal L}_{\gamma\gamma}}{dz}=2z\int_{z^2/x_{max}}^{x_{max}}
 \frac{dx}{x} f_{\gamma/e}(x)f_{\gamma/e}(z^2/x),
\end{eqnarray}

where $f_{\gamma/e}$ is the photon structure function, the fraction of the energy of the incident electron carried
by the back-scattered photon $x$, which are interfaced by the CompAZ code \cite{Zarnecki:2002qr}. At low $x$ part ($x\leq0.1$), the photon spectrum is not properly described and underestimated, and it is qualitatively better for larger values of fraction $x$ of the longitudinal momentum of the electron beam. However, for $x > 2(1+\sqrt{2})\simeq 4.8$, the high energy photons can disappear through $e^+e^-$ pair creation in its collision with a following laser photon.

The Feynman Rules are extracted by the program FeynRules \cite{FeynRules2.0} from the Lagrangian of IDM, then outputs to Universal FeynRules Output(UFO) files \cite{UFO}. For the cross-section calculation and simulation for signal and backgrounds, we make use of the Monte Carlo event generator MadGraph@NLO(MG5)\cite{MadGraph5}. PYTHIA6 \cite{Pythia} is utilized for parton shower and hadronization with the options of ISR and RSR included. Delphes \cite{Delphes} is then employed to account for the detector simulations and MadAnalysis5 for analysis, where the (mis-)tagging efficiencies and fake rates are assumed to be their default values in Delphes. The IDM mediator width is automatically computed by using the MadWidth module for each parameter point.

\begin{figure}
\begin{center}
\includegraphics[width=0.49 \textwidth]{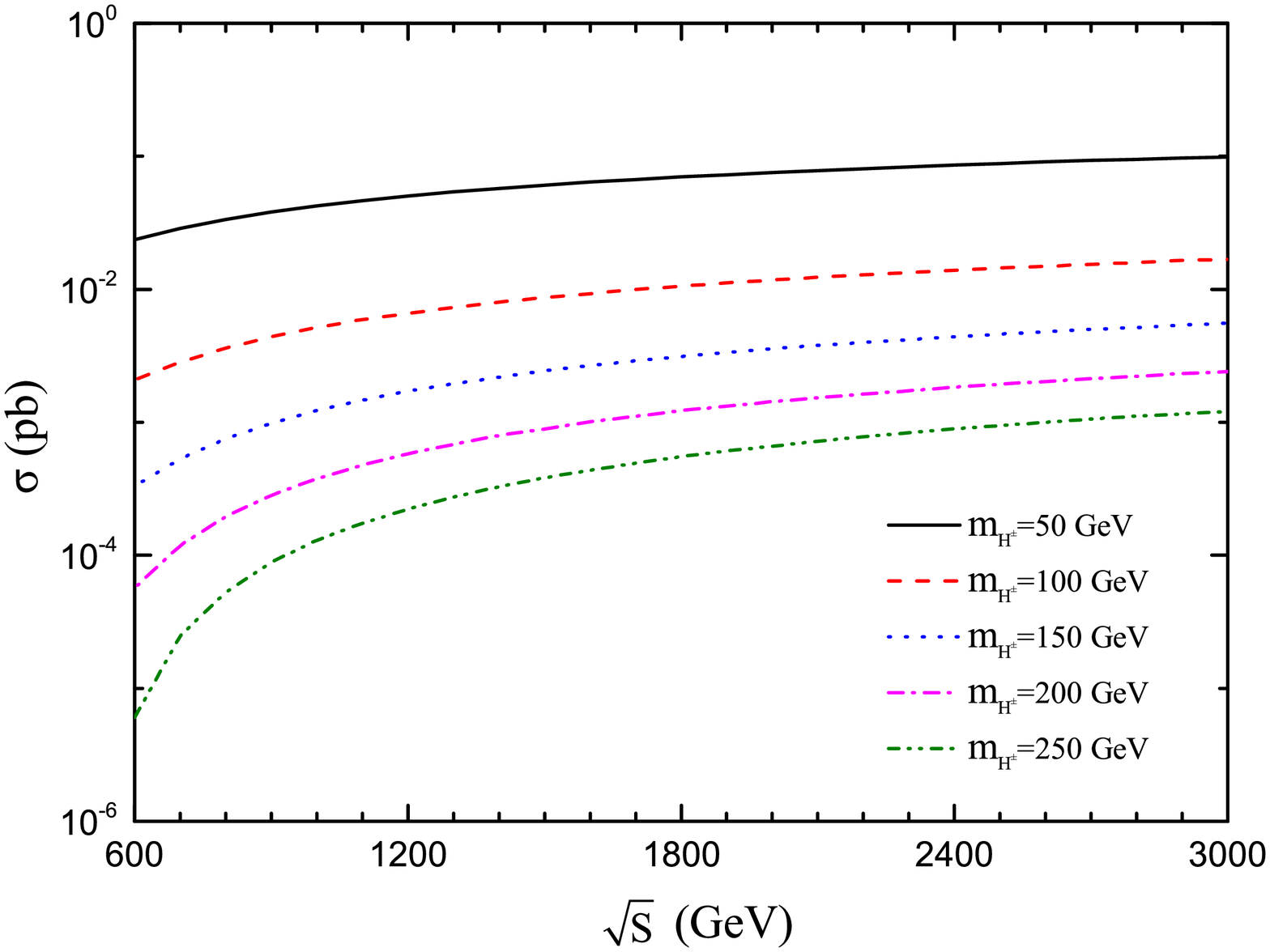}
\includegraphics[width=0.49 \textwidth]{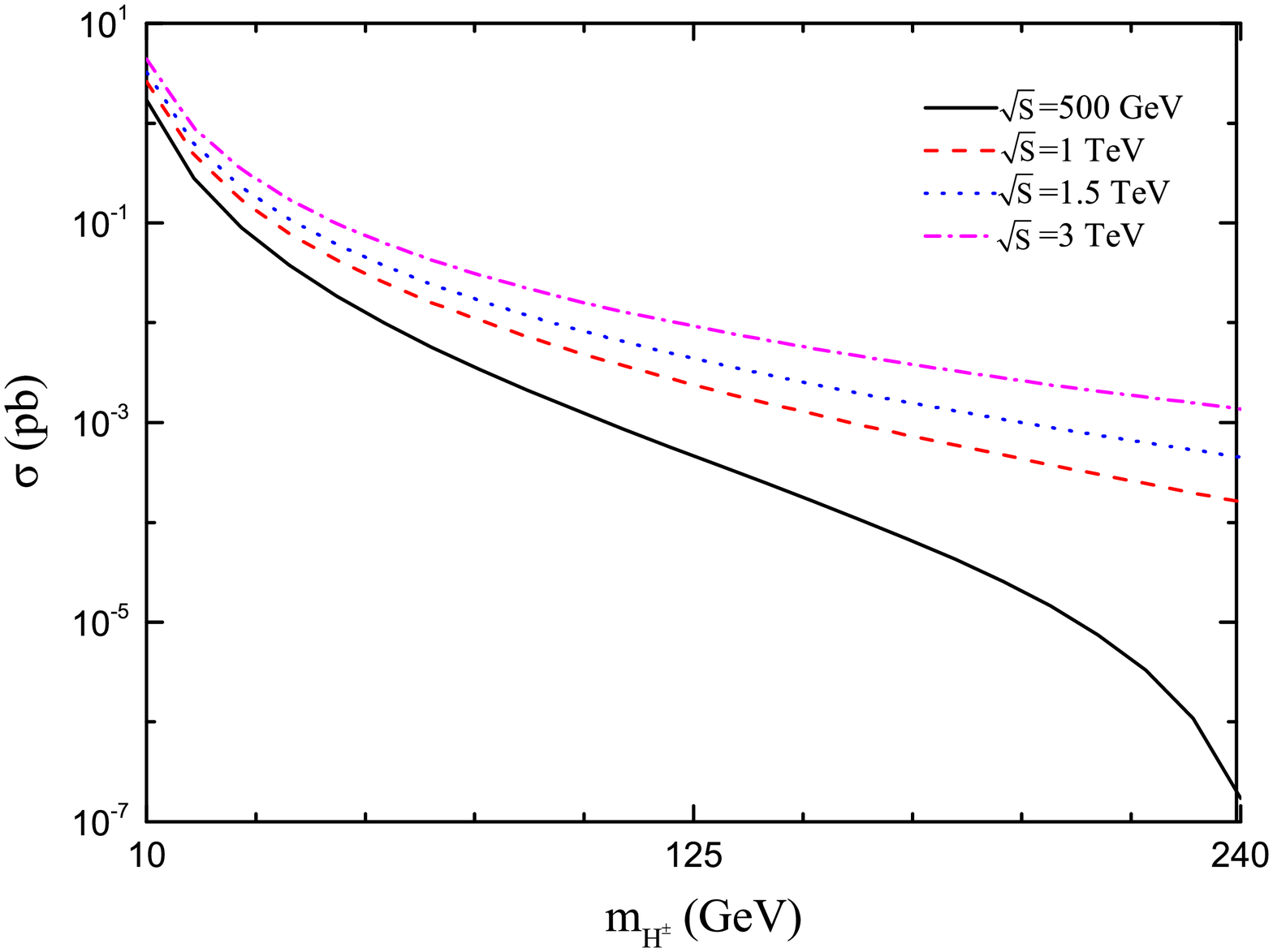}
\caption{ \label{squre} (color online) The cross sections of $e^+e^- \to \gamma \gamma \to H^+ H^-$ production as a function of the center-mass energy
when the mass $m_H^{\pm}$ is fixed(left) and the cross sections as functions
of the mass $m_H^{\pm}$ when the center-mass energy is fixed (right) in IDM at $\gamma\gamma$ collider.}
\end{center}
\end{figure}

\par
In Fig.\ref{squre} (left), we present the cross sections as functions of the
colliding energy $\sqrt{s}$ for process $e^+e^- \to \gamma \gamma \to H^+ H^-$ by taking
$m_H^{\pm} = 50,100,150,200,250~ {\rm GeV}$, separately. From this figure, we can see that,
with the increment of the colliding energy $\sqrt{s}$, the total cross section for the process
$e^+e^- \to \gamma \gamma \to H^+ H^-$  increases rapidly at first. When the colliding energy $\sqrt{s}$ reaches about 1 TeV,
the total cross section increases slightly. Consequently, we can obtain larger cross section
for process $e^+e^- \to \gamma \gamma \to H^+ H^-$ by raising the colliding energy
$\sqrt{s}$. In Fig.\ref{squre} (right), the total cross section is plotted for different mass of $m_H^{\pm}$ at
$e^+ e^-$ collider by taking $\sqrt{s}$ = 500, 1000, 1500 and 3000 GeV. With the increment of charge Higgs mass $m_H^{\pm}$,
the total cross section is decreasing. When its mass is close to a half of the centre of mass energy, the cross section quickly is approaching zero.

\vskip 5mm
\section{Signal and Background}

\begin{figure}
\begin{center}
\includegraphics[width=0.9 \textwidth]{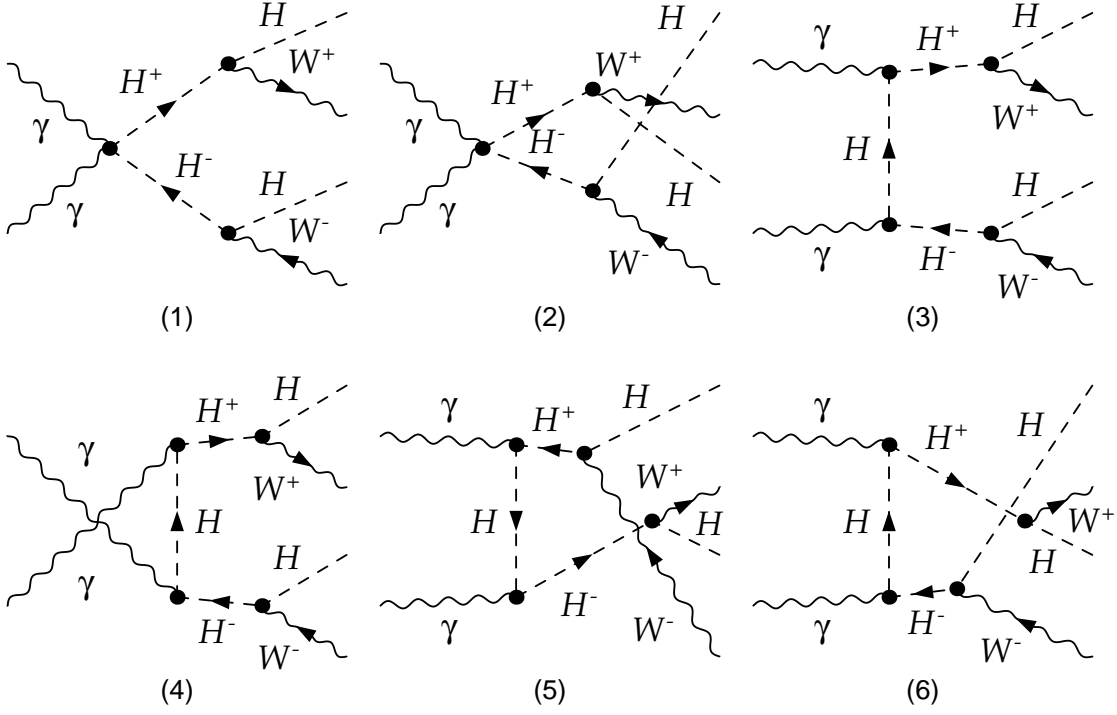}
\caption{ \label{fig5} The Feynman diagrams for process $\gamma \gamma   \to H^{+} H ^{-} \to W^{+} W^{-} H H $ in IDM .}
\end{center}	
\end{figure}

Since the lightest scalar boson $H$ is stable, the charge Higgs $H^{\pm}$ particles will eventually decay into the $H$ and SM particles.
In this section, we perform the Monte Carlo simulation and explore the sensitivity in photon-photon collider through the following channel,
\begin{eqnarray}
\label{process2}
\gamma \gamma   \to H^{+} H ^{-} \to W^{+} W^{-}  H H,
\end{eqnarray}
$H$ assumed as the lightest scalar particle in the IDM leave missing energy in detector and make it almost impossible to reconstruct events.
$W$ boson decay to a electron or muon and its antineutrino. The Feynman diagrams for the process $\gamma \gamma \to H^{+} H ^{-} \to W^{+} W^{-}  H H$ are presented in Fig.\ref{fig5}.

The dominant signal for pure leptonic channel is $\ell^{+} \ell^{-}+\slashed{E}_{T}$  in the IDM, where $\ell$  = {e, $\mu$}, which can obtain from  either $H^{\pm} \to W^{\pm} H$, with $W^{\pm} \to \ell^{\pm} \nu $ or  $H^{\pm} \to W^{\pm} A$, with $W^{\pm} \to \ell^{\pm} \nu, A \to H Z, Z \to  \nu  \overline {\nu}$, depend on the choice of parameters. The contribution from second decay chain can be neglect comparing the first case.
Thus, we will focus on the process $\gamma \gamma  \to H^{+} H ^{-} \to W^{+} W^{-} H H $, with the decay $W^{\pm} \to \ell^{\pm} \nu $.
The cross sections of the production  $\gamma \gamma  \to H^{+} H ^{-}\to W^{+} W^{-}  H H$ in the IDM with $\sqrt{s}=500$ GeV for the benchmark points are given in table \ref{cross}.

\begin{table}[htb]
	\begin{center}
		\caption{\label{BP} The cross sections for the process $e^+e^- \to \gamma \gamma \to H^{+} H^{-}\to W^{+} W^{-} H H$ in the IDM with $\sqrt{s}=500$ GeV. \label{cross}}
		\vspace{0.2cm}
		\begin{tabular}{|c|c|c|c|c|c|c|c|}
			\hline
			& BP1& BP2 & BP3 & BP4 &BP5 &BP6  \\ \hline
			$\sigma(fb)$&1.362&1.014 &4.318&3.296&3.683&5.168\\ \hline
			
		\end{tabular}
	\end{center}
\end{table}

In the pure leptonic channel, the signal of this process is two leptons $l^+l^-$ plus missing $E_{T}$, and the main backgrounds of the Standard Model are mainly $W^{+} W^{-}$ , Drell-Yan process, top-quark pair production ($t\bar{t}$), $WZ$, and $ZZ$ processes. For the Drell-Yan process, the two leptons are always back-to-back, and the missing $E_{T}$ is very small, which can be easily distinguished from the large missing $E_{T}$ signal.
The final state of top-quark pair production contains a large number of hadrons, which can also be well eliminated in the photon photon collider. $WZ$ and $ZZ$ processes can also suppressed seriously
by the two leptons invariant mass cut of $Z$ boson. These backgrounds can be neglected after suitable cuts. Therefore, we will not list the these backgrounds in the following analysis.
We will analysis the main irreducible background $W^{+} W^{-}$ production.

In our simulation, we first employ some basic cuts for the selection of events:
\begin{eqnarray}
p_T^\ell > 20 ~GeV,  ~|\eta_\ell| < 2.0, ~\Delta R_{\ell\ell} > 0.4,
\end{eqnarray}
where $p_T^{\ell}$ and $\eta_{\ell}$ are the transverse momentum and the pseudorapidity of the
leptons. $\Delta R = \sqrt{\Delta \phi^2 + \Delta\eta^2}$ is the particle separation among the leptons
in the final state with $\Delta \phi$ and $\Delta\eta$ being the separation in the azimuth angle and rapidity.
The $\eta_{\ell}$ acceptance region avoids the gap between barrel and endcap,
where the misidentification probability is the highest.

\begin{figure}
	\begin{center}
		\includegraphics[width=0.45 \textwidth]{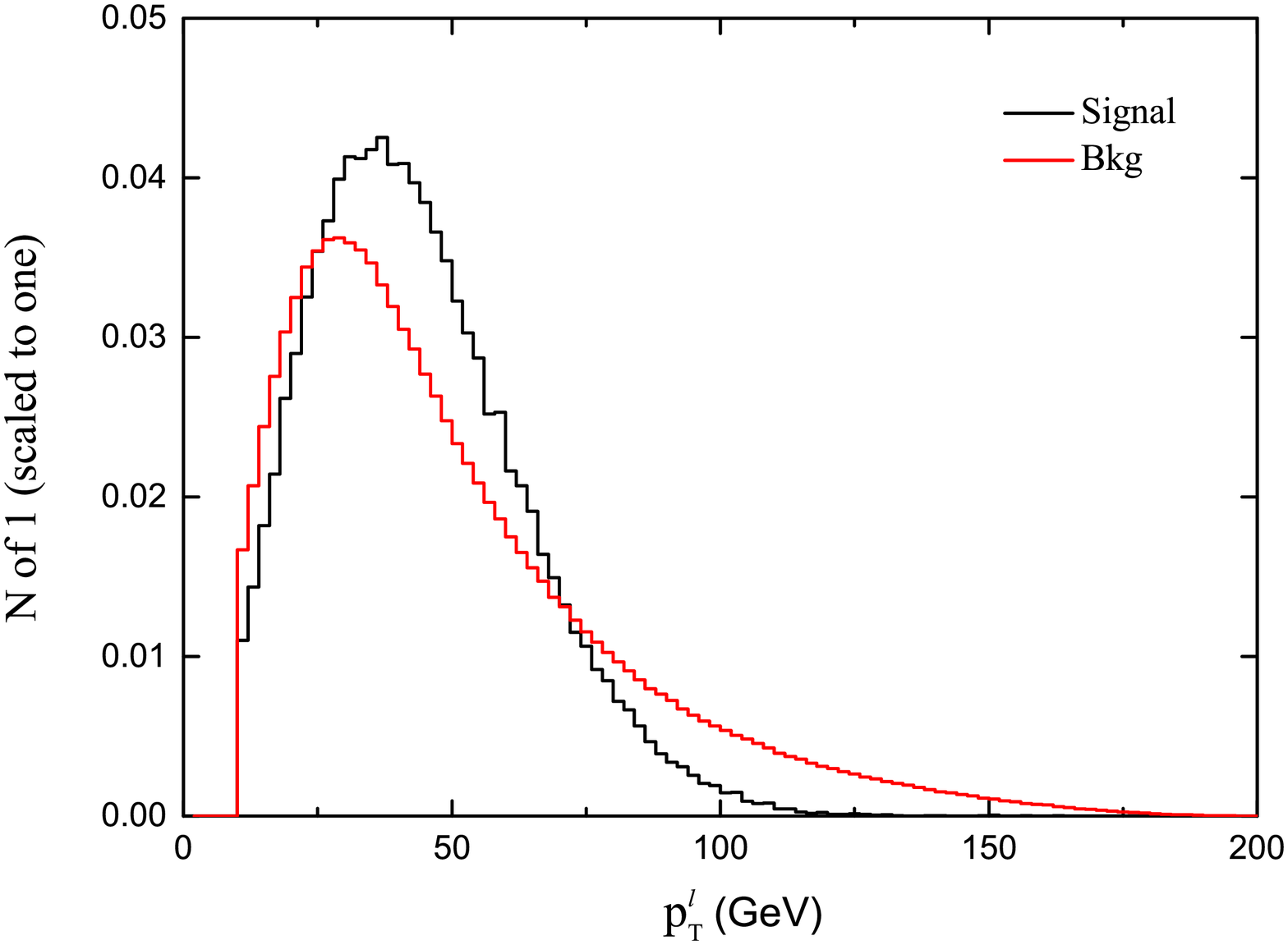}
		\includegraphics[width=0.45 \textwidth]{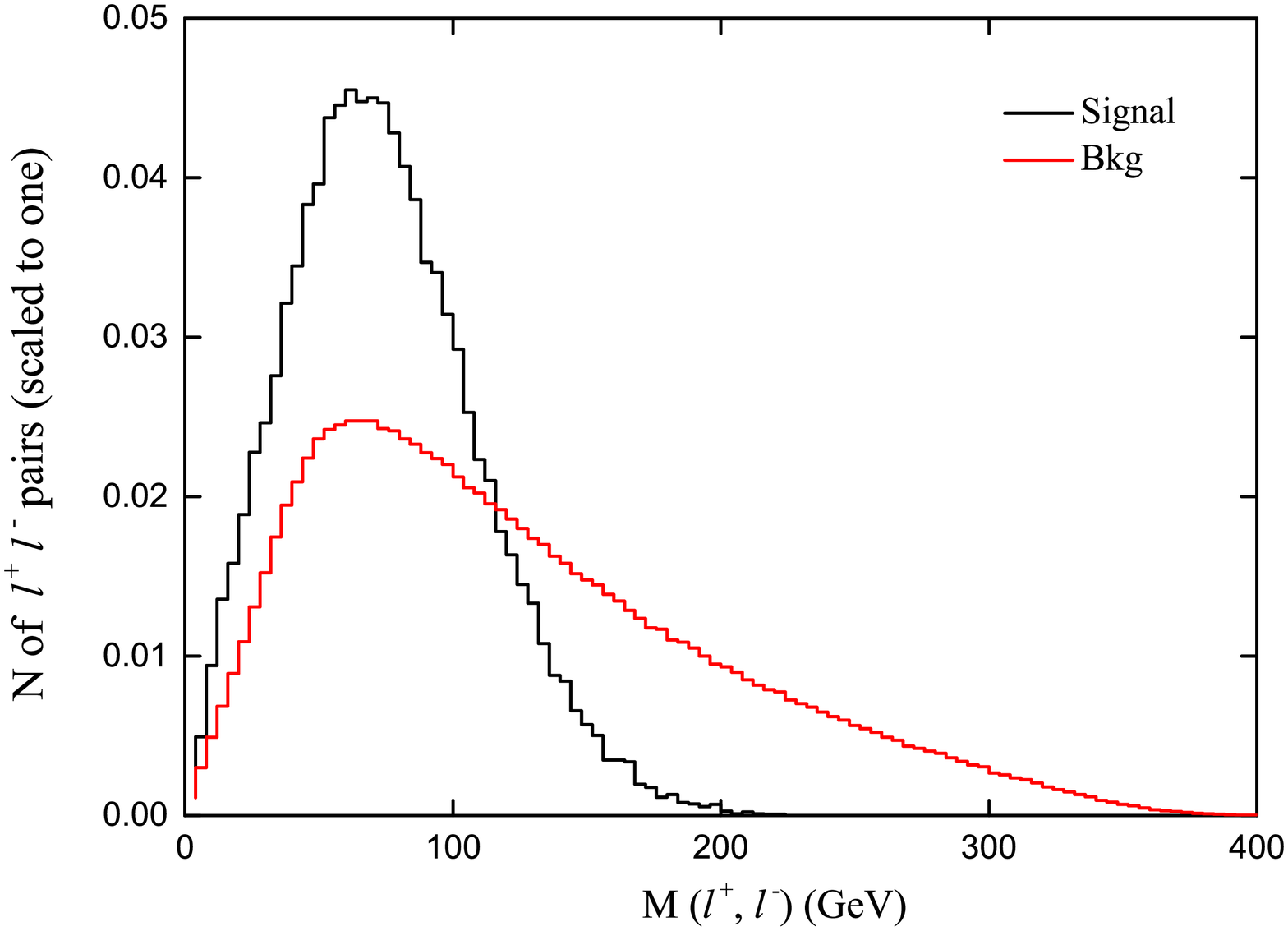}
		\includegraphics[width=0.45 \textwidth]{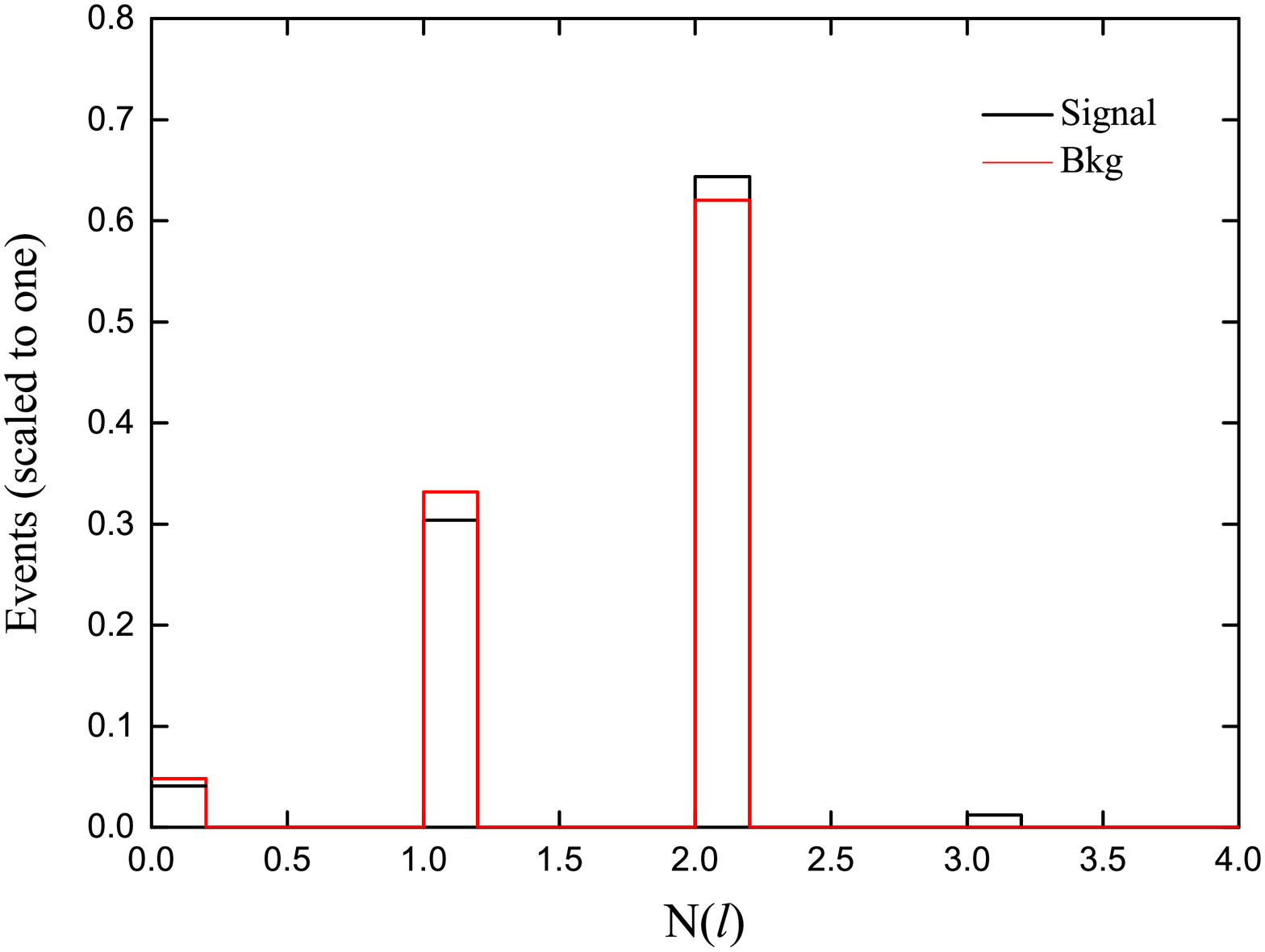}
		\includegraphics[width=0.45 \textwidth]{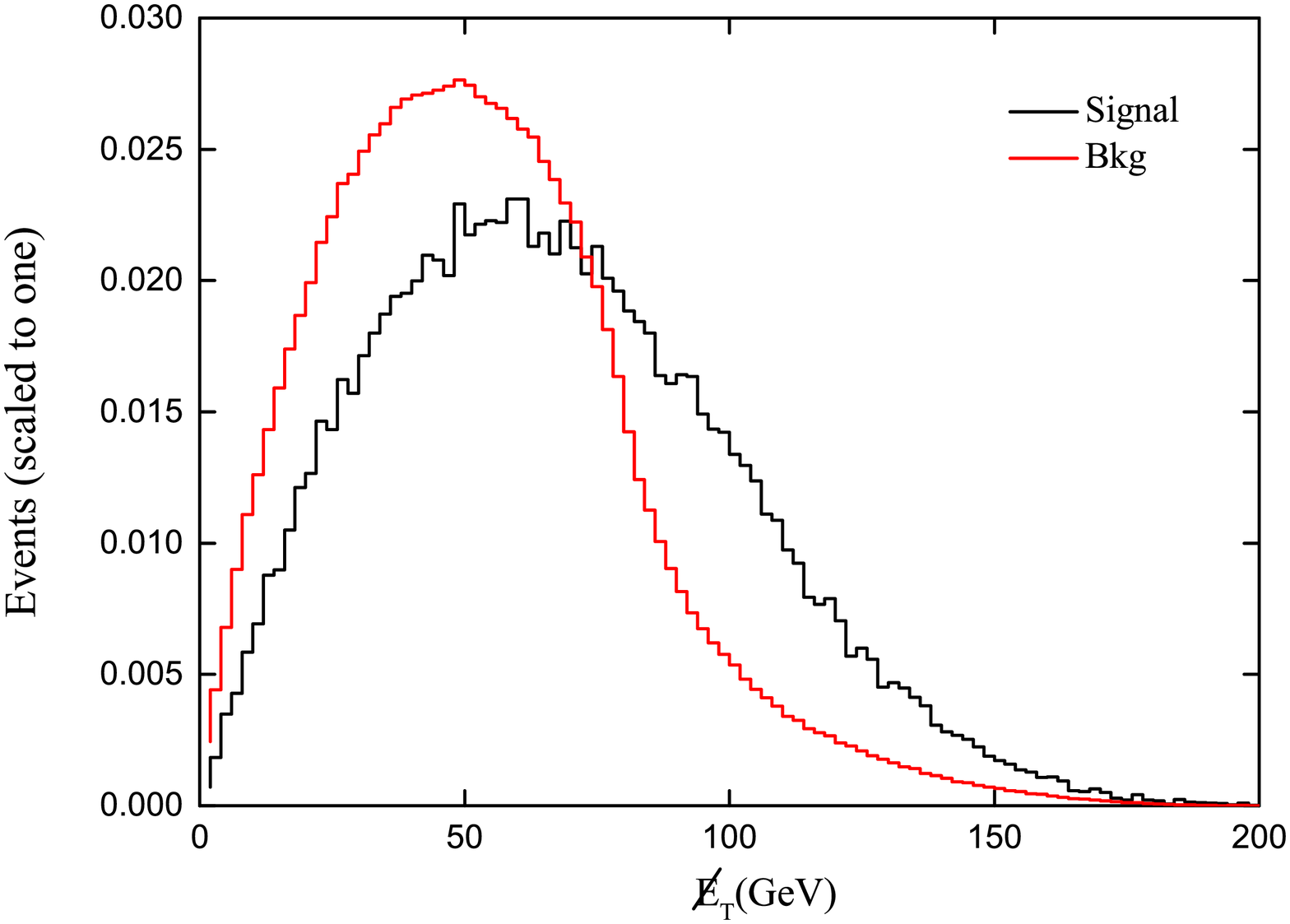}
		\includegraphics[width=0.45 \textwidth]{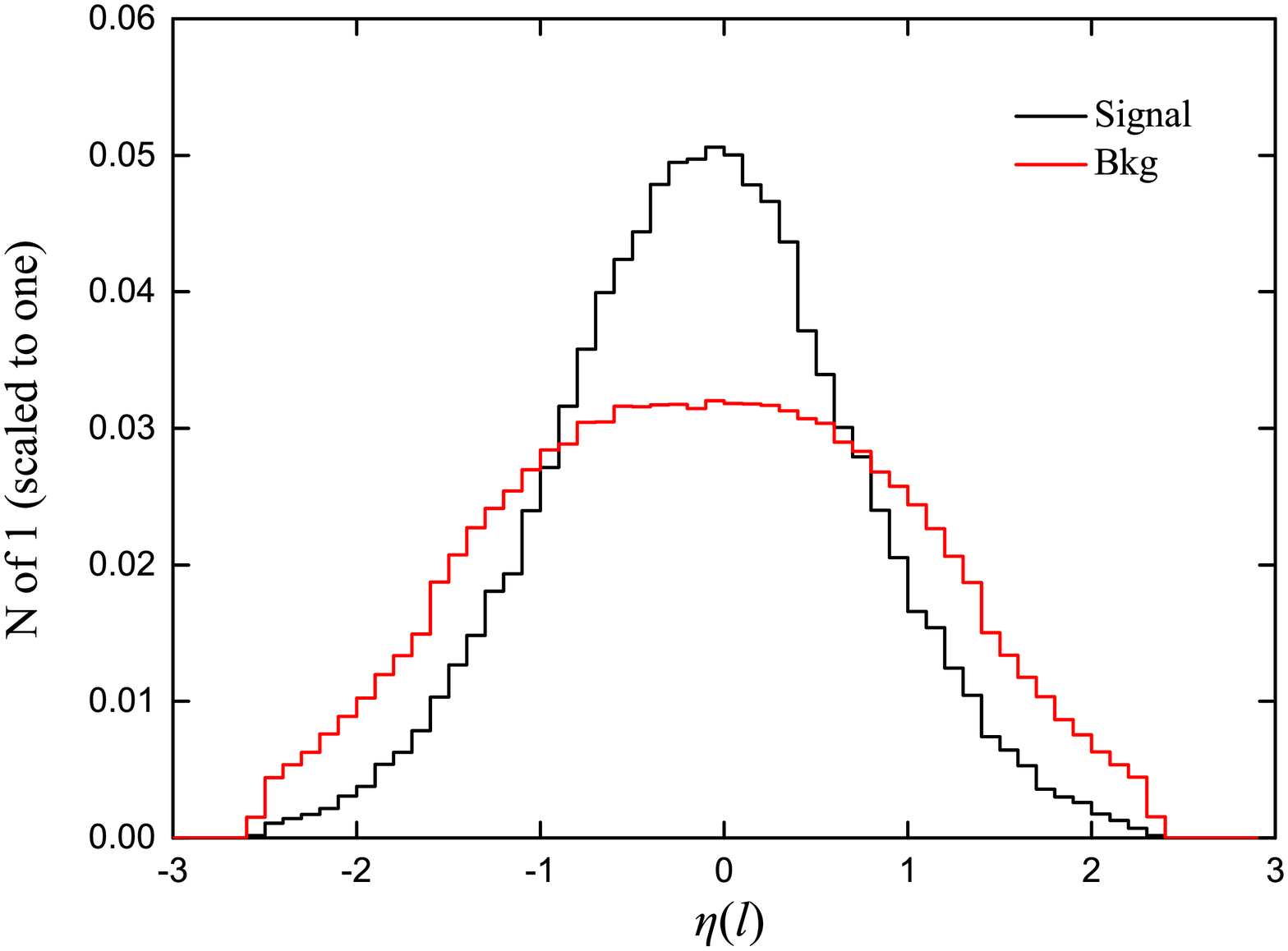}
		\includegraphics[width=0.45 \textwidth]{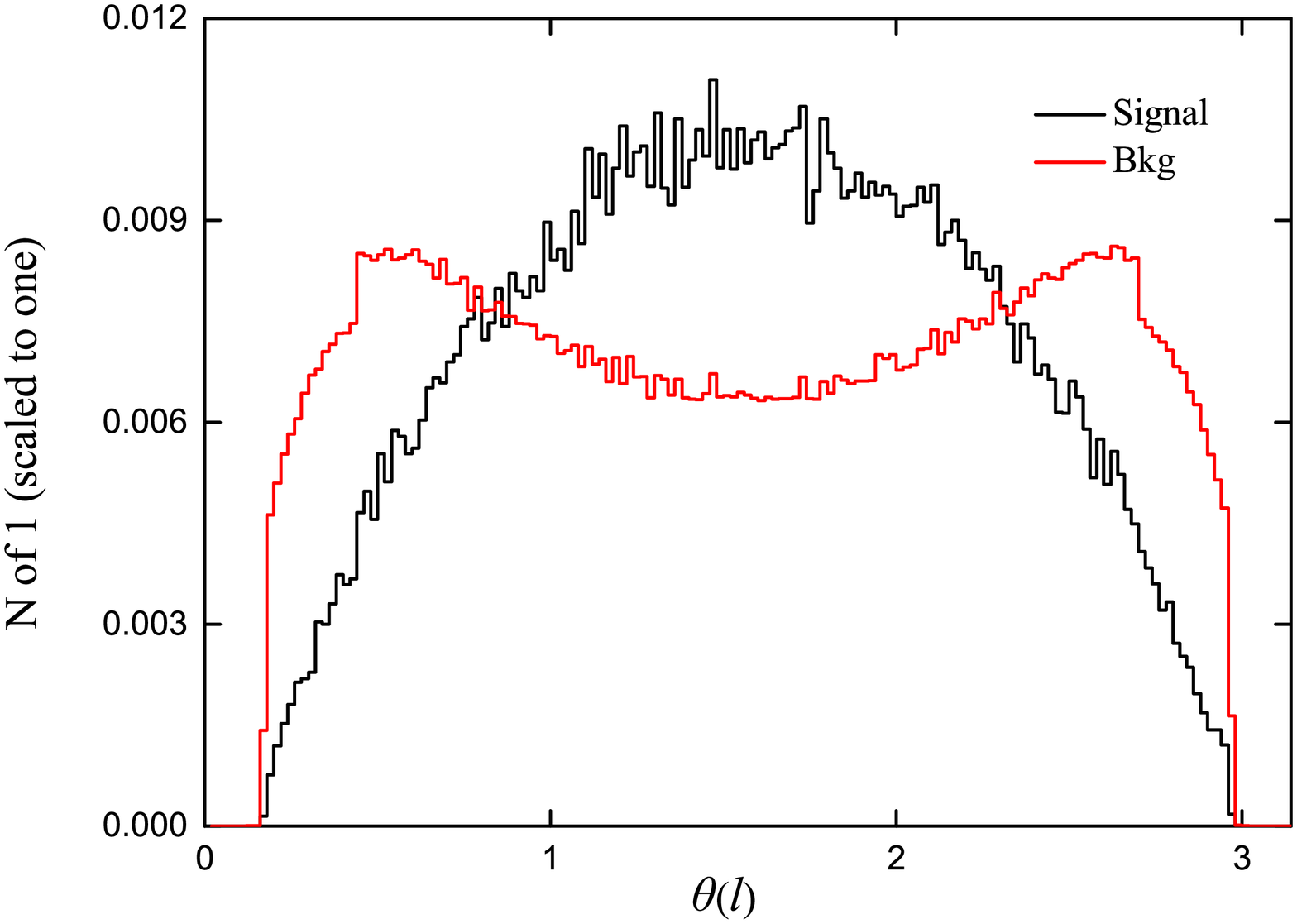}
		\caption{ \label{fig7} (color online) Normalized distributions of  the leptonic transverse momentum $p_{T}^{\ell}$, the invariant mass  $M(\ell^+,\ell^-) $, numbers of lepton $N(\ell)$, the transverse missing energy $\slashed E_{T}$ , angle $\theta$, pseudorapidity $\eta$  for the signal and background with $\sqrt{s}$ =500 GeV.}
	\end{center}	
\end{figure}

According to the differential distribution between the signal and background, we can improve the ratio of signal to background by making suitable kinematical cuts. In Fig.\ref{fig7},
we show the distributions of some kinematical variables for the signal and background at 500 GeV. We first select $N(\ell) =  2$ , the signal almost concentrate low $p_{T}^{\ell}$ region,
so we reject  $p_{T}^{\ell} > $ 70 GeV. Then, because of signal decrease faster than background in high invariant mass region, $M(\ell^+,\ell^-) < 125 ~\rm {GeV}$ is required.
Finally, we require the transverse missing energy $\slashed E_{T} > 95 $ GeV  to improve the discovery significance.

For a short summary, we list all the cut-based selections here:
\begin{itemize}
	\item[(1)] Basic cut: $p_T^{\ell}>20$ GeV, $|\eta_{\ell}|<2.0$
	and $\Delta R_{\ell\ell}> 0.4$;
	\item[(2)] Cut 1 means the basic cuts plus requiring select $ N(\ell) =2 $ ;
	\item[(3)] Cut 2 means Cut 1 plus requiring $p_T^{\ell} < 70~\rm {GeV}$;
	\item[(4)] Cut 3 means Cut 2 plus requiring the invariant mass of two leptons $M(\ell^+, \ell^-) < 125~\rm {GeV}$ ;
	 \item[(5)] Cut 4 means Cut 3 plus requiring the transverse missing energy $\slashed E_{T} >  95~\rm{GeV}$.
\end{itemize}

\begin{table}[htb]
	\begin{center}
		\caption{ The number of events for the signal ( $W^{+} W^{-} H H$ ) in BP2 and main backgrounds ( $W^{+} W^{-}$) after the cut flows are listed in the brackets at the 500 GeV with integrated luminosity $L = 3000 fb^{-1}$.  The values of discovery significance $S/\sqrt{B+S}$ at each step of cut are also shown. \label{cutflow}}
	{
\begin{tabular}{|c|c|c|c|c|c|c|}
	\hline
	Cuts & Signal&Background&$S/\sqrt{B+S}$  \\ \hline
	Basic cuts &$3.039\times10^{3}$&$2.616\times10^{6}$&1.880 \\ \hline
	Cut 1&$1.956\times10^{3}$&$1.623\times10^{6}$&1.534\\ \hline
	Cut 2 &$1.592\times10^{3}$&$1.036\times10^{6}$&1.563\\ \hline
	Cut 3&$1.528\times10^{3}$&$7.294\times10^{5}$&1.787\\ \hline
	Cut 4&$3.438\times10^{2}$&$2.904\times10^{2}$&13.653\\ \hline
\end{tabular} }
\end{center}
\end{table}

The results of the number for the signal in BP2 and backgrounds (with luminosity $= 3000 fb^{-1}$) are shown in Table \ref{cutflow} at each step of the cuts. The values of the discovery significance $S/\sqrt{B+S}$ are also shown, where $S$ and $B$ are the numbers of signal and total background events, respectively.
After applying several cuts, the background can be reduced greatly, the discovery significance $S/\sqrt{B+S}$ can reach $13.653\sigma$. Thus, we have potential for observing the IDM effect though the charged Higgs $H^{\pm} $ pair in some parameter space with large luminosity at the $ \gamma\gamma $ collider.

\begin{figure}
	\begin{center}
		\includegraphics[width=0.8\textwidth]{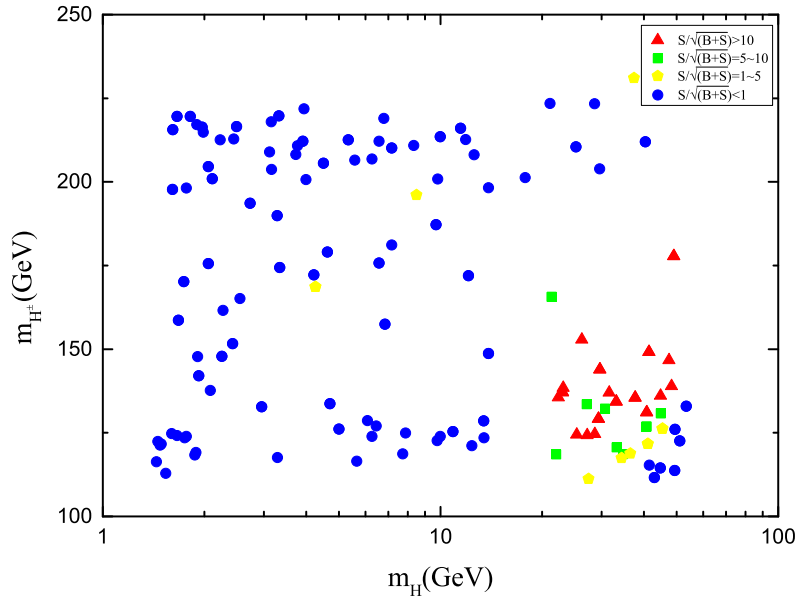}
		\caption{ \label{fig8} (color online) The signal background ratio at different reference points in  $m_H-m_{H^\pm} $ plane.}
	\end{center}	
\end{figure}

In Fig.\ref{fig8}, we present the distribution of the parameter point for the discovery significance $S/\sqrt{B+S}$ in the plane $m_{H^{+}}-m_H$ with the integrated luminosity of $3000fb^{-1}$ at $\sqrt{s}$ = 500 GeV. The parameter points with different colour represent the value of the significance. We investigate the effects of coupling parameter $\lambda_2$, $\lambda_L$ and the scalar even particle mass $m_A$  and
find that the cross section has little change when varying these input parameters. From Fig.\ref{fig8}, we find that the parameter points with high significance are mainly concentrated in the range of $m_H$ from 10 to 50 GeV and $m_{H^{+}}$ from 110 to 180 GeV. If the CEPC or ILC can be built, these parameter points in the IDM model has potential to be detected or excluded.

\vskip 5mm
\section{Summary}
The Inert Doublet Model is one of the most simple extension of the Standard Model, which provide a scalar DM particle candidate. In this paper, we have investigated the double charged Higgs $ H^{\pm} $ pair production in IDM at the $ \gamma\gamma $ collider. Assuming that the lightest scalar Higgs is the dark matter particle, we have calculated the corresponding relic abundance, scanned the IDM parameter space, and obtained the parameter points satisfying the relic abundance of dark matter in our universe. We analyzed the pure lepton decay process of the double charged Higgs $H^{\pm} $ and the backgrounds of the Standard Model, and optimised the selection criteria employing suitable cuts on the kinematic variables to maximise the signal significance. We found that with high luminosity option of the  $ \gamma\gamma $ collider, this channel has the potential to probe the IDM in the mass range of 1-250 GeV. In a scenario with light dark matter of mass about 10-50 GeV, charged Higgs in the mass range of around  110-180 GeV provides the best possibility with a signal significance of about  $10  \sigma $  at an integrated luminosity of about 3000 $fb^{-1}$.

\section{Acknowledgments}
This work was supported by the National Natural Science Foundation of China (No.11205003, No.11305001, No.11575002, No.11935001).

%-------------------------------------------------------------------------------------------------------

\end{document}